\documentclass[12pt,a4paper]{article}
\usepackage{latexsym,amsfonts,amssymb,fleqn,euscript}
\usepackage{url,matstyle}
\usepackage{color}
\usepackage{graphicx}
\definecolor{red,green,blue}{rgb}{0.5,0.5,0.7}
\definecolor{cyan,magenta,yellow,black}{cmyk}{0.5,0.5,0.5,1}

\def\R2n{{\bf {R}}^{2n}}

\def\R{{\bf {R}}}

\newtheorem{Figure}{Figure}[section]

\title{The GMc-interpretation of Quantum Mechanics}
\author{Christian ~Jansson}

\begin{document}

\maketitle

\vspace*{0.5cm} Institute for Reliable Computing, Technical
University Hamburg--Harburg, Schwarzenbergstra{\ss}e 95, 21071
Hamburg, Germany, e-mail: jansson@tu-harburg.de, Fax: ++49 40
428782489.

\begin{center}
\vspace*{0.5cm} \noindent{\bf{Work in progress}}
\end{center}

\begin{abstract} The GMc-interpretation (Gravitation-Motion of mass-light with maximal speed c)
is a consistent approach to quantum mechanics very closely related to classical physics. Several postulates are formulated that are satisfied in classical physics, general relativity
theory, quantum theory and thermodynamics. In this interpretation, particles are always particles, never
waves, and many paradoxes can be easily understood or avoided. In particular, the
postulates allow to explain the measurement problem and the concept of interaction-free
measurements; the latter allows to find objects without ``touching'' them, and sometimes the
phrase ``seeing in the dark'' is used. Additionally, the concepts of complementarity,
uncertainty, decoherence, locality and realism are investigated. One key property of this
interpretation is that all observed probabilities and interference patterns are known before
a particle is in the experiment. Ontological questions are discussed.
\end{abstract}

\section{Introduction} The measurement problem is viewed very differently in
quantum mechanics. Despite the considerable philosophical
differences, however, the different interpretations almost agree on the practical question of
what results from a quantum laboratory measurement. Very tricky are
interaction-free measurements, a type of measurement in quantum
mechanics that detects the position or the state of an object without any
interaction occurring between the measurement apparatus and the
object. Examples include certain double-cavity optical systems and
the Elitzur-Vaidman bomb-testing problem \cite{EliVai93}. These examples are
verified by experiments. In this context, Vaidman writes in \cite{Vaidman} that
if a neutron, while being inside an interferometer, can feel being in two places
and moving in two different directions simultaneously, it must have
schizophrenic experiences.

Several different interpretations are given for the paradoxes that
occur in quantum theory. For example, Schr{\"o}dinger viewed the
wavefunction associated to a particle as a density smeared out over
the space. Max Born interpreted it as the probability distribution
for its position. The Copenhagen interpretation with two dynamics
(Schr{\"o}dinger's equation and the collapse postulate) is the most
popular among scientists. This interpretation rejects questions like
``where is the particle before measuring it'' as meaningless.
Decoherence together with many worlds or many minds is also very
popular nowadays. The difference when interpreting the wavefunction is that now the world must split, in
contrast to other interpretations where the particle has to split in
form of waves. Several other interpretations exist that have more or less
paradoxes. Many physicists, however, consider non-instrumental
questions (in particular ontological questions) to be irrelevant to
physics. They fall back on David Mermin's expression ``Shut up and
calculate''.

An agnostic view is due to von Weizs{\"a}cker \cite{Cramer}, who, while
participating in a colloquium at Cambridge, denied that the
Copenhagen interpretation asserted: "What cannot be observed does
not exist". Instead, he suggested that the Copenhagen interpretation
follows the principle: ``What is observed certainly exists; about
what is not observed we are still free to make suitable assumptions.
We use that freedom to avoid paradoxes.''

In the following I describe the GMc-interpretation as
according to von Weizs{\"a}ckers opinion. The underlying concepts
consist of clear formulations distinguishing future, present and past events,
together with some basic
postulates that are satisfied in physics, and which are formulated
in an absolutely ordered space of higher dimension. This universe also
describes potential events that may take place in the present.
Exemplary quantum mechanical experiments are explained in detail
within the GMc-interpretation . There, particles always remain
particles, waves are not required, only one universe is needed,
the splitting of worlds and minds is avoided,
measurements can be described in terms of classical mechanics, only one dynamics suffice, and
the quantum probabilities are known before a particle is in the
experiment. Questions do not have to be rejected, but can be answered in
terms of appropriate postulates. This paper, moreover, also serves
to ´specify statements presented in \cite{Janssona}.


\section{A Thought Experiment}
During the last years I have given short courses in quantum computation to students studying electrical engineering and computer science. Most of these students have an intuitive understanding of classical mechanics and electrodynamics, but the postulates in quantum mechanics, the mathematical formalism, and the interpretations appear very confusing to them (and also to me). In order to become more familiar with this formalism we made some thought experiments in its broadest usage, i.e. of employing imaginary situations to help us understand the way things really are.

One experiment, for instance, was to imagine at our universe from the outside as follows: Let's stop our universe for one moment, only for one moment. Everything is at rest, the particles, the macroscopic objects, the speed, the entropy, just everything. It's like a photo. Then let's look at a certain electron or at a fullerene. We have stopped the time and the entropy, nothing happens, but there is this small particle. It is not measured, since we are looking from the outside, there is no measurement apparatus. It does not collapse, because the time is at rest. Who believes that this small particle you can see is a wave and goes all paths in this moment?

My students do not believe in coherent superpositions. Their arguments are like: We have observed in very many experiments, may be millions, that we can prepare single photons, electrons and other microscopic objects. All these objects arrived always as the same objects. We have never observed a wave. This suggests to believe what we see, and not what we don't see. Is this true?

What is true? What is reliable? What is correct? What then shall we believe?  Mirman \cite[page~434]{Mirm06} writes:

\begin{quote}
Is the literature correct? We are all prisoners of what has been published - all of us, authors, readers and other scientists. Much of what is in the literature is wrong. Much of science that has been long known to be certainly true is now known to be certainly false. ... Ultimately the surest course is to read, think, and be skeptical. Each of us is responsible for our own beliefs. Ones based on careful thought are the ones we should respect the most.
\end{quote}

I would like to add that paradoxes are a strong sign of incredibility, and each pursuable theory avoiding paradoxes and providing realism is more reliable.

I tried to find an interpretation capturing the point of view of looking at the universe from the outside. It turns out that this interpretation is very closely related to classical mechanics, from the point of view of understanding as well as from the mathematical framework. I am very grateful for any comments and suggestions on this subject.

\section{The GMc-Interpretation}
Physics describes our universe mainly within the two frameworks relativity theory and quantum
theory. A basic quantity in physics is the relative time, which is defined
in terms of particles, namely the photons that present us photos from the past. The relative time is
used to describe all kinds of interactions and the four-dimensional curved space-time with
all the matter and the particles. The relative time is a concept of the present,
and has no meaning for the future or the past. In fact, this time is not absolute and  takes different values
depending on the reference frames. Thus, it is an dependent observable.

The GMc-interpretation strictly separates past, present and future.
The future gives us freedom to choose when there are alternatives. This freedom or flexibility can be described by more independent variables or, in other words, by a higher dimension.
In the GMc-interpretation, the four-dimensional space-time, which exists only at present, is embedded in an absolutely ordered higher dimensional space describing all possible events. There, velocity vectors are free for use and not tied to the relative time. The same holds true for other quantities like polarization or spin.

A {\it classical state} is a notion that fully describes the considered property of a particle or a system. It can be represented in a linear space.  If we are only interested in gravitation, mass, and motion, a classical state is defined by position and velocity. In the GMc-interpretation these quantities are represented in a six-dimensional {\it ordered universe}:

(1) There is a three-dimensional space of {\it points} (representing positions), say $X$.

(2) There is a three-dimensional space of {\it vectors} (representing velocities), say $V$.

(3) There is an {\it absolute time $T$} defining an ordering of the universe (which should not be confused with the relative time).

(4) The universe is described by the family $(X,V,T)$ with classical states $(x,v,T)$, and the ordering of classical states $(x_1,v_1,T_1) \le (x_2,v_2,T_2)$ if $T_1 \le T_2$.

A classical state describes the degree of freedom which a particle has in the future when the velocity is not tied to positions by the relative time, and $T$ describes an ordering where you cannot go back. The ordering may be viewed as the entropy. Since all particles only live in our present four-dimensional universe, the above ordered universe $(X,V,T)$ provides an outside frame that the present universe traverses. A slightly different imagining is to view this family as a film reel, and the present universe is represented as the photos for different ordering times $T$.

At present, the freedom of particles is reduced if one of the independent variables or quantities is expressed by another. In this case we call the variables {\it entangled}. They are called {\it complementary} if additionally these quantities complete to a state.  Since the velocities $v_i$ can be expressed by the corresponding position coordinates $x_i$ for $i=1,2,3$ and the relative time, these variables are entangled, and in this case we say time-entangled. Moreover, they are complementary. The variables $v_j$ and $x_i$ are not time-entangled for $i \neq j$.  The kinetic energy is time-entangled with the relative time itself. In the present universe, everything is realized, and time-entangled variables reduce the number of independent variables and hence the dimension.

In \cite{Mirm06} a proof is given that our present universe would be impossible in any dimension but 3+1. It is astonishing that a change of any number in any of the formulas by even 1 would make any dimension, thus any universe, impossible. That the universe allows and has galaxies, stars, planets, even life, thinking life, that all the conflicting conditions do not conflict and are met, is beyond stunning. This deep awareness, however, does not contradict the existence of an higher dimensional cover.

The above family $(X,V,T)$ can also be used to describe other situations. Properties like spin or polarization, for example, can be handled In the same framework.

For the ordered universe  $(X,V,T)$ we claim the following postulates:

\begin{quote}

\emph{Postulate 1: (Ordered Universe)} The physical laws of the ordered universe must be compatible with the laws of our present four-dimensional universe.
\end{quote}
The laws of the ordered universe in some sense provide a ``metaphysics''
beyond the present physical world. An interpretation of the ordering $T$ is the entropy , sometimes called an arrow of time.  Hence, in classical mechanics, a state $(x(t),v(t))$ in the present universe yields immediately the state $(x,v,T)$  at ordering time $T$ , and vice versa. Compatible means that this transformation does not lead to any contradictions to or violations of true physical laws formulated for the present universe.

One of the most spectacular consequences of quantum mechanics has been the development of a new kind of logic, called quantum logic. This modification
of classical logic was first proposed by Birkhoff and von Neumann and is still
subject of debate and controversy. In classical logic any two propositions
simultaneously have a sharp truth-value; that is, they are either both true, or both false, or one is
true and the other one false. Precisely this assumption is denied by quantum mechanics.
As a consequence, for example, the proposition (x and y) or (x and not-y), being equal to x classically,
is no longer true in quantum logic, because x and y may not simultaneously have a sharp
truth-value. Therefore, quantum logic provides a probabilistic interpretation. Here, we presuppose
\begin{quote}

\emph{Postulate 2: (Classical Logic)} The ordered universe $(X,V,T)$ and the universe of the present satisfies the rules of classical logic.
\end{quote}

In this interpretation, quantum logic has only a meaning in the past, and comprises the loss of information about non-measured and unknown events of the present.

An observable describes alternative classical states or properties for a particle. For each state there is assigned an observable value that is a function of position and velocity, like the energy or momentum. We only consider the finite dimensional case. The generalization to the infinite dimensional case is straightforward. Mathematically, there are different ways to describe observables and properties, which can be transmitted as usual. In the following, the approach and the notion mainly used in quantum computing is presented. There, an observable with its alternative properties is described by an orthonormal system of basis vectors ($|r_i\rangle$) in the $n$-dimensional complex Hilbert space, with observable values $r_i$. In the GMc-interpretation we claim for the present and the future:
\begin{quote}

\emph{Postulate 3: (Particles)} For a given observable, particles are always exactly in one state $|r_i\rangle$.
\end{quote}

Hence, a particle can never be in two states at the same ordering time, and therefore it is explicitly forbidden that particles have wave-like properties.  In the future, it is not clear in which state particles and systems will be. In order to make predictions possible, we define a {\it probability state} w.r.t. an observable by
\begin{equation}\label{eq1*}
| \Psi, T \rangle := | \Psi \rangle = \sum\limits_{i=1}^m \psi_i |r_i\rangle, \; \mbox{where} \sum\limits_{i=1}^m |\psi_i|^2 = 1.
\end{equation}
The interpretation of a probability state is: {\it If a particle is in the future (say at ordering time $T$) in one of the basis states, then it is in state $|r_i\rangle$ with probability $|\psi_i|^2$.} This definition includes also the classical state, where a particle is in state $|r_i\rangle$ with probability one. It excludes, however, the case where a particle is in two states at the same ordering time.

When performing several experiments with particles in the same probability state $| \Psi \rangle$, i.e. we consider an {\it ensemble}, we obtain for the observable values the average value
\begin{equation}\label{eq1a*}
\langle R \rangle = \sum\limits_
{i=1}^m r_i |\psi_i|^2 = \langle \Psi| R | \Psi \rangle,
\end{equation}
where $R$ denotes the Hermitian matrix $R = \sum\limits_{i=1}^m r_i |r_i\rangle \langle r_i |$.

Next, we consider a general dynamics which holds true for all present and future events.
\begin{quote}
\emph{Postulate 4: (Dynamics)} Probability states are transformed linearly into probability states for increasing $T$.
\end{quote}

This dynamics contains unitary transformations as well as projections with normalization. It is exactly one dynamics where the linear transformations are applied to probability states and not to coherent waves. The latter is the reason why a collapse does not occur although projections are allowed. The dynamics, for example, can be applied when building up an experiment. This irreversible process, expressed by increasing ordering $T$, linearly changes the probability states.
Let

\begin{equation}\label{eq2*}
|\Psi \rangle = \sum\limits_{i=1}^m \psi_i|r_i\rangle \quad \mbox{and} \quad |\varphi\rangle = \sum\limits_{j=1}^n \varphi_j |s_j\rangle
\end{equation}
be two probability states, which can be transformed linearly into one another, then there exists a linear operator $U$ such that
\begin{equation}\label{eq3*}
U |r_i \rangle = \sum\limits_{j=1}^n U_{ij} |s_j\rangle \quad \mbox{for} \quad i=1, \ldots, m.
\end{equation}
Therefore, if a particle has property $|r_i\rangle$, then later it will have property $|s_j\rangle$  with probability $|U_{ij}|^2$. It is not required that $U$ is unitary or invertible. Moreover, for the probability states the linearity implies
\begin{equation} \label{eq4*}
U|\Psi \rangle = \sum\limits_{j=1}^n \left(\sum\limits_{i=1}^m U_{ij} \psi_i \right)|s_j\rangle , \quad \varphi_j = \sum\limits_{i=1}^m U_{ij} \psi_i, \quad \mbox{for} \quad j = 1, \ldots, m.
\end{equation}
The tensor product is used for describing composite systems of particles.
\begin{quote}
\emph{Postulate 5: (Systems of particles)} The probability state of a composite system of particles is the tensor product of the probability states of each particle.
\end{quote}
{\it Product probability states} are as usual defined as states where the coefficients are the result of multiplying all coefficients of the probability states for the single particle. All other probability states in the tensor product contain dependencies, and are called {\it entangled}. In principle, entanglement describes probabilistic dependencies between various particles.

\section{Other Physical Theories}
Classical mechanics satisfies the previous postulates. The motion of  particles is described
by the Euler-Lagrange equations in the four-dimensional space resulting in classical
states in the six-dimensional ordered universe. Classical logic is assumed in classical
mechanics, and there are always classical states as presumed in Postulate 3. Since position
and velocity are not tied by time and can be chosen arbitrarily, the simple linear
transformations ``displacement of the position''  and ``rotation and contraction of velocity vectors'' are
sufficient for fulfilling the Euler-Lagrange equations in the present universe. Hence, the classical dynamics
satisfies Postulate 4. Macroscopic objects, from projectiles to parts of machinery, as well
as astronomical objects, such as planets, stars and galaxies, can be described very easily
as entangled classical or probability states. In a macroscopic solid object, for example, two neighboring
atoms have almost the same position and velocity, and the degree of entanglement is almost
maximal. The fact that the moon surrounds the earth can also be described by states entangled as
macroscopic objects and entangled with the relative time. Here, we must have in mind that
the relative time is an observable. The degree of entanglement is much weaker than in the
first example.

Thermodynamics can be viewed as the bridge between macroscopic and microscopic properties of
systems. In the statistical approach all macroscopic properties (temperature, volume,
pressure, energy, entropy, etc.) are derived from the properties of moving particles and the
interactions between them. Therefore, the postulates are satisfied, and thermodynamics can
be viewed mainly as a theory of entanglement.

In relativity theory  the events  in space-time coordinates $(x(t),v(t),t)$ depend on the reference frame, and the Lorentz transformation converts between two different frames, where one observer is in constant motion with respect to the other one. Frequently, it is mentioned that the principle of relativity implies that there is no absolutely reference frame. This is correct when only looking at reference frames from the point of view of relativity. On the other hand, the Lorentz transformation defines an equivalence relation. Hence, at fixed ordering time $T$, we take the equivalence class of an event, denoted by $(x,v,T)$, as the object of an absolutely ordered universe. Hence, relativity theory can be embedded in $(X,V,T)$, and with the same argument as above
the  postulates are fulfilled.

Quantum mechanics satisfies the above postulates with unitary transformations and projectors.
Postulate 3, however, implies that only probability states are allowed, not coherent
wave-superpositions. In other words, this means that decoherent states are completely
sufficient to describe the future and the present. This approach is very similar to the many worlds or
many minds interpretation, but in GMc the splitting of worlds or minds is avoided.

{\it Observe that these physical theories can be embedded into the postulates because of the higher dimension, and thereout the resulting freedom to choose independently quantities from one another that are time-entangled in the present. }

Many physicists believe that there are some simple physical laws that underlie our universe, and these laws are unique in the sense that no extra constants are demanded that determine  the theory. This lucid attitude is due to Einstein \cite[page 63]{Einstein69} written at his age of sixty-seven:

\begin{quote}
I would like to state a theorem which at present can not be based upon anything
more than upon a faith in the simplicity, i.e., intelligibility, of nature: there are
no arbitrary constants ... that is to say, nature is so constituted that it is possible
logically to lay down such strongly determined laws that within these laws
only rationally completely determined constants occur (not constants, therefore,
whose numerical value could be changed without destroying the theory).
Over the course of the twentieth century, that program has worked remarkably well
\end{quote}

In relativity theory, the maximal speed of light can set to one when measuring lengths in feet and time in nanoseconds. In this sense $c$ is a rational constant.

The contents of the previous set of simple postulates is that we live inside an absolutely ordered higher-dimensional universe that can be described without arbitrary constants, that is completely linear, that satisfies the rules of classical logic, that can be described by real properties, and in this universe particles stay over as particles without any wave-like properties. Of course, this is not the complete set of simple laws, but in the following it is shown that they are sufficient and close-fitting to describe interaction-free measurements without any paradoxes. Moreover, several ontological arguments for these postulates will be discussed.

\section{Experiments} In this section we consider  interferometer-type experiments exemplary. The mathematical formulation is straightforward, but the interpretation  is different and leads to an understanding of interaction-free measurements. Very similar, other experiments like photon polarization or Stern-Gerlach devices can be treated.

\begin{table}
\begin{center}
\includegraphics[height=5cm]{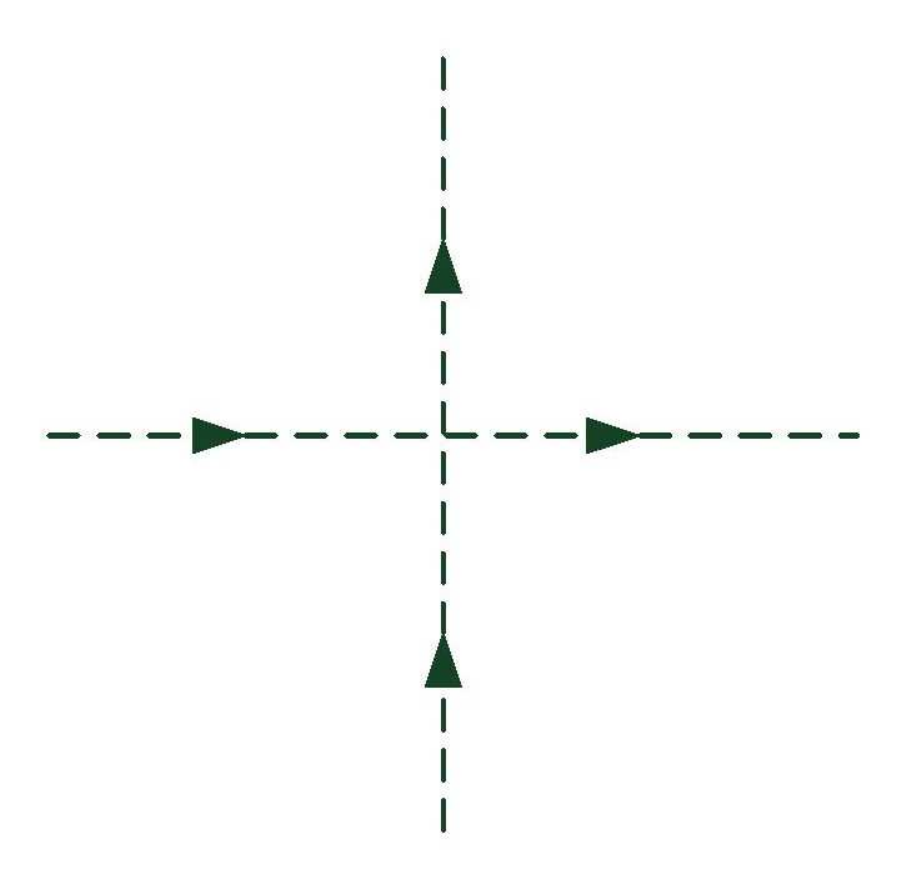}

\begin{Figure} \label{fig0}
A cross.
\end{Figure}
\end{center}
\end{table}

Experiments with interferometers are composed of several devices where the particles move on straight lines. We describe and explain the interference patterns by applying the previous postulates to these devices; particles are not in the experiment. Firstly, there are crosses (see Figure \ref{fig0}) where particles can move. A cross has the property that a particle can either move horizontally  or vertically, which is described by the two classical states ${1 \choose 0}$ and ${0 \choose 1}$, respectively. Mathematically, the cross can be expressed by the two-dimensional identity matrix, because
\begin{equation} \label{eq1}
\left(\begin{array}{cc} 1 & 0 \\ 0 & 1 \end{array}\right) {1 \choose 0} = {1 \choose 0},
\left(\begin{array}{cc} 1 & 0 \\ 0 & 1 \end{array}\right) {0 \choose 1} = {0 \choose 1}.
\end{equation}

The probability state
\begin{equation} \label{eq2}
|\psi \rangle = \alpha {1 \choose 0} + \beta {0 \choose 1}, \quad |\alpha|^2 + |\beta|^2 = 1
\end{equation}
has a simple physical meaning: {\it If a particle will be put in the cross, then it will move with probability $|\alpha|^2$ horizontally, and with probability $|\beta|^2 $ it will move vertically}.  In the following experiments we consider the special case where $\alpha$ and $\beta $ are equal to $\frac{1}{\sqrt{2}}$. The switch  to general amplitudes is straightforward.

The next device is a reflector (see Figure \ref{fig1*}), and its  interpretation is:  {\it If a particle will be put into the experiment, and if it will move horizontally, then it is rotated and will move vertically. If it is coming from below, then it will be rotated and will move horizontally}. The reflector is given by
\begin{equation} \label{eq2+}
R = \left(\begin{array}{cc} 0 & i \\ i & 0 \end{array}\right), \quad R{1 \choose 0} = i{0 \choose 1}, \quad R{0 \choose 1} = i {1 \choose 0}.
\end{equation}

\begin{table}
\begin{center}
\includegraphics[height=5cm]{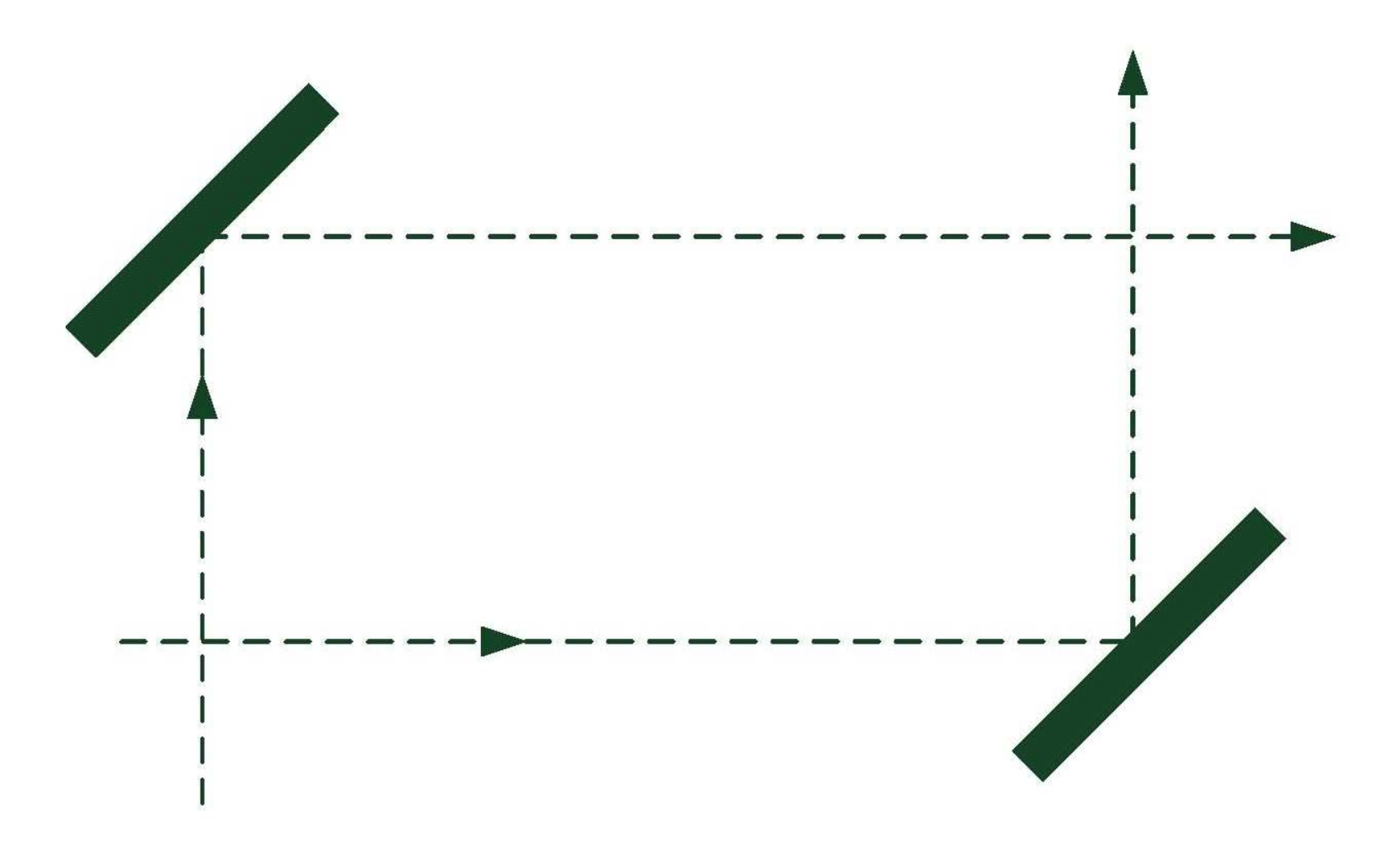}

\begin{Figure} \label{fig1*}
Two reflecting mirrors.
\end{Figure}
\end{center}
\end{table}

A rotation requires an acceleration. Since the velocities and the positions can be chosen free and are not entangled by time, forces and accelerations, important for the present universe, have no meaning for future prognosis.
We use complex numbers for convenience, but everything could also be expressed by real rotations or density matrices.

A half-silvered mirror, or more general a beam-splitter (see Figure \ref{fig1}), is a device $H$ with two lines where particle can either pass the half-silvered  mirror  with probability $1/2$, or can be reflected with the same probability. A reflection is a rotation of the velocity about $90^{\circ}$ that can be realized by multiplying with the complex number $i$. This device is represented mathematically by
\begin{equation} \label{eq1+}
H = \frac{1}{\sqrt{2}} \left(\begin{array}{cc} 1 & i \\ i & 1 \end{array}\right).
\end{equation}

\begin{table}
\begin{center}
\includegraphics[height=6cm]{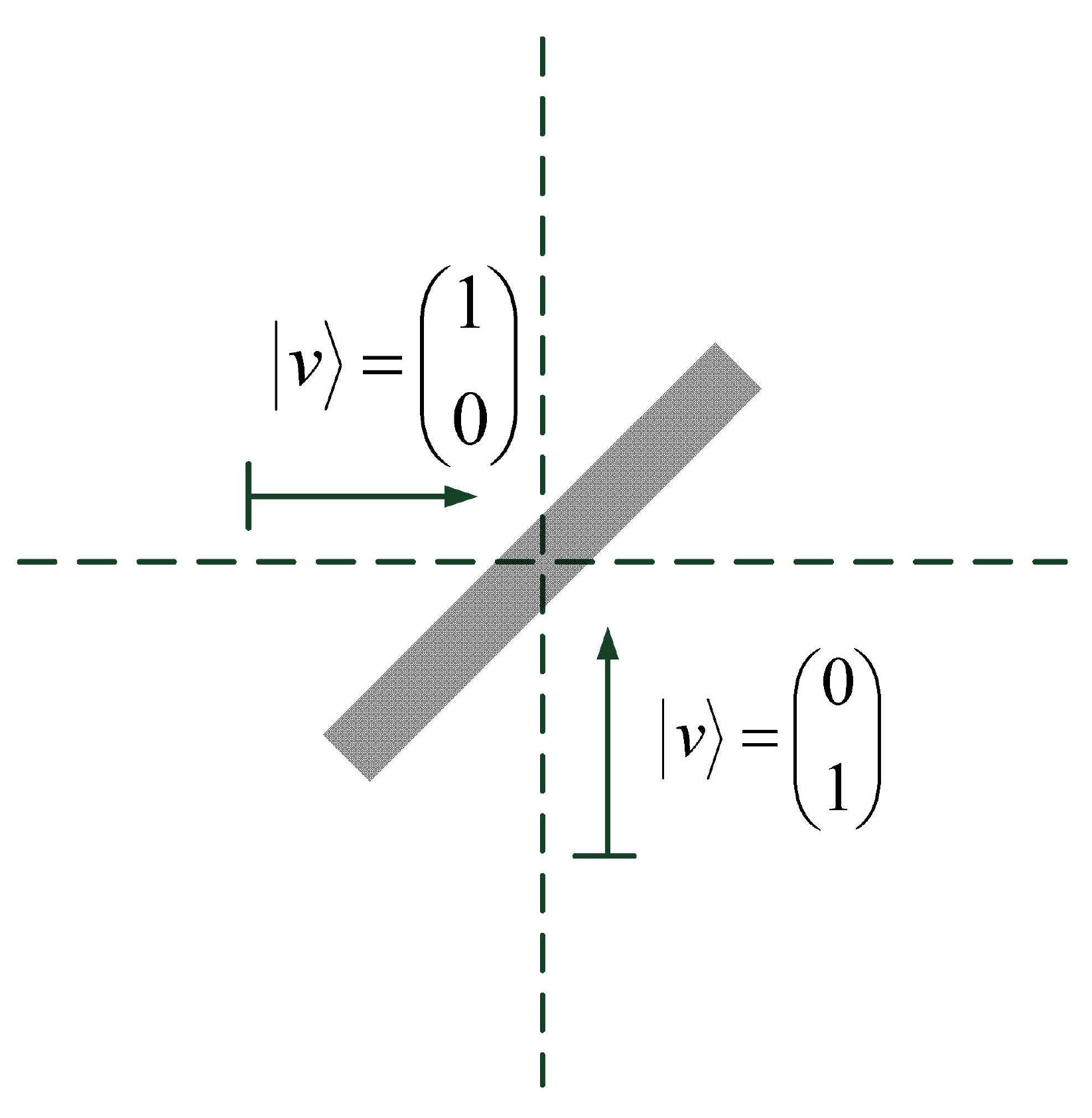}
\end{center}

(1) \hspace*{1cm}
\begin{minipage}[c]{6cm}
\includegraphics[height=5cm]{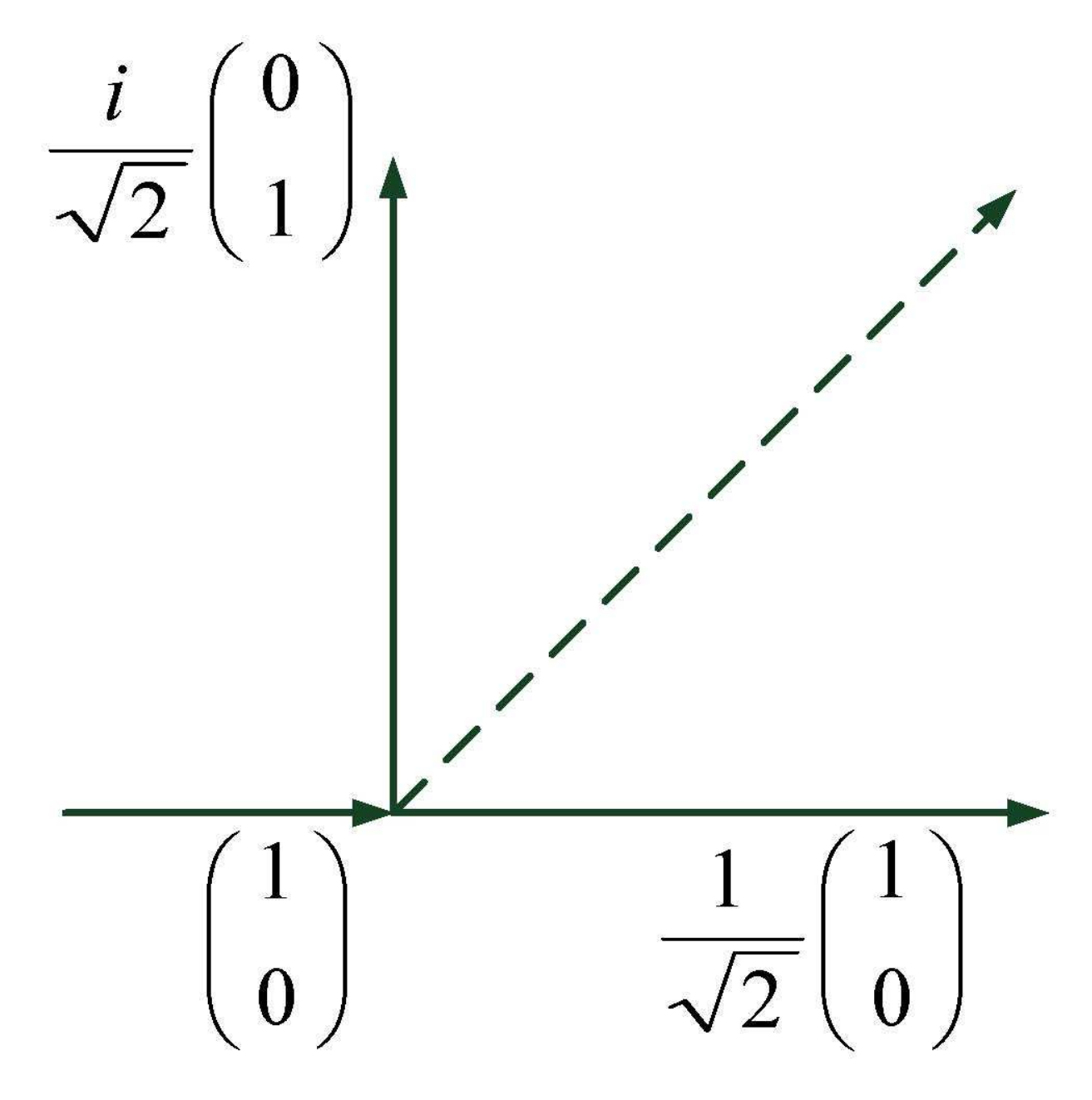}
\end{minipage}
  \hspace*{2cm}
\begin{minipage}[c]{6cm}$H{\displaystyle {1 \choose 0}} = \frac{1}{\sqrt{2}} {1 \choose 0} + \frac{i}{\sqrt{2}}{0 \choose 1}$
\end{minipage}

(2) \hspace*{1cm}
\begin{minipage}[c]{6cm}
\includegraphics[height=5.2cm]{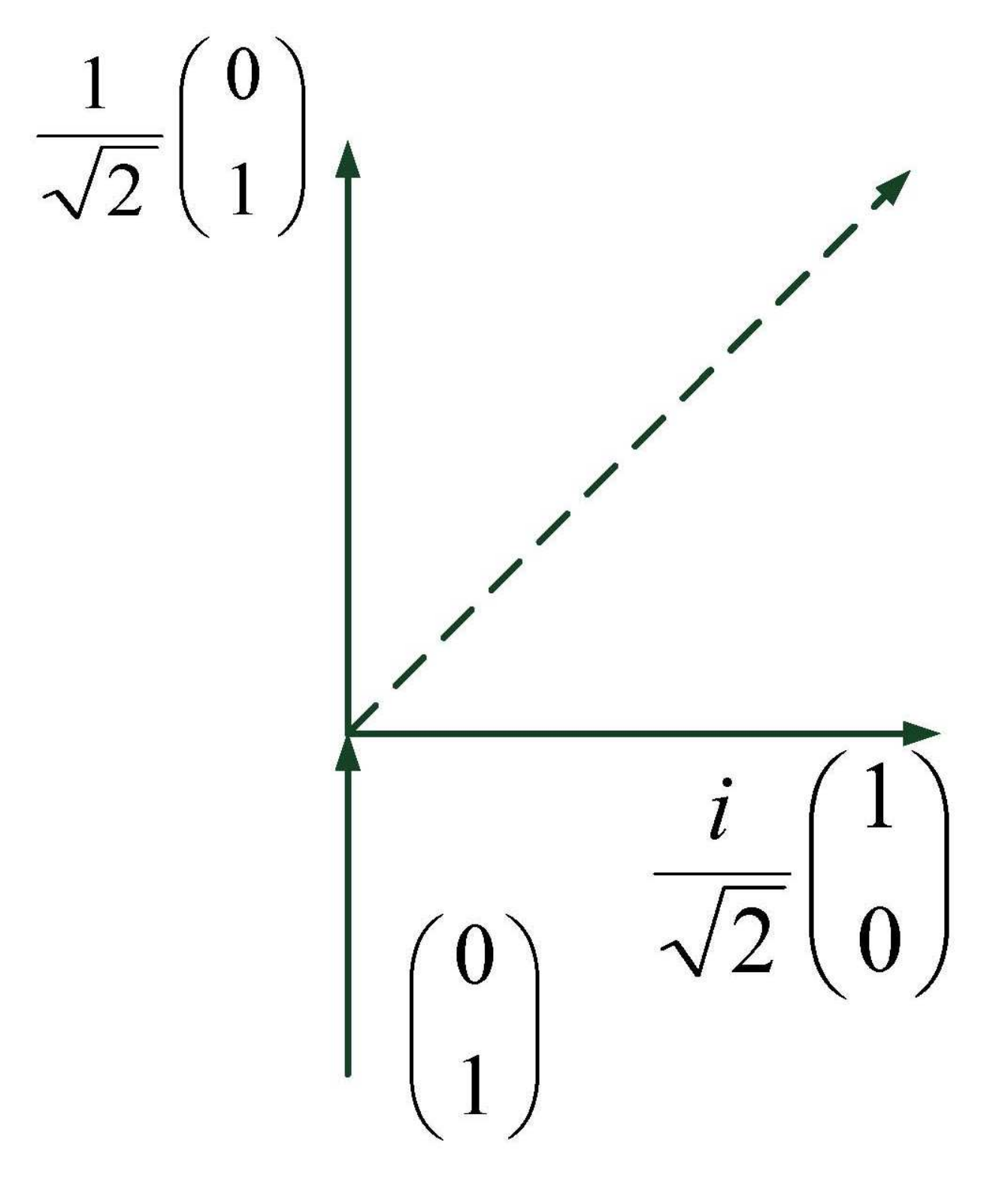}
\end{minipage}  
\hspace*{2cm}
\begin{minipage}[c]{6cm}$H{\displaystyle {0 \choose 1}} = \frac{i}{\sqrt{2}} {1 \choose 0} + \frac{1}{\sqrt{2}}{0 \choose 1}$
\end{minipage}

\begin{center}
\begin{Figure} \label{fig1}
A half-silvered mirror $H$ with rules.
\end{Figure}
\end{center}
\end{table}

Firstly, the experiment described in Figure \ref{fig2} is discussed. This set up is the composition of the devices $H, R, P_1,P_2,$ where the last two operators are the projectors describing both detectors, and the two-dimensional identity matrix, which is suppressed for convenience. The composition is linear, and for the lower path to detector I we obtain the probability state
\begin{equation}\label{eq3+}
\begin{array}{lcl}
{\displaystyle P_1 R H{1 \choose 0}} & = & {\displaystyle P_1R \left(\frac{1}{\sqrt{2}}{1 \choose 0}+ \frac{i}{\sqrt{2}}{0 \choose 1}\right)}\\[0.5cm]
 & = & {\displaystyle P_1 \left(\frac{i}{\sqrt{2}}{0 \choose 1}- \frac{1}{\sqrt{2}}{1 \choose 0}\right)}\\[0.5cm]
 & = & {\displaystyle \frac{i}{\sqrt{2}}{0 \choose 1}}.
\end{array}
\end{equation}
Analogously, the upper path to detector II yields
\begin{equation} \label{eq4+}
P_2 RH {1 \choose 0} = -\frac{1}{\sqrt{2}}{1 \choose 0}.
\end{equation}
Summarizing, the particle will be observed with probability $\frac{1}{2} = |\frac{i}{\sqrt{2}}|^2$ by detector I and with the same probability by detector II.

\begin{table}
\begin{center}
\includegraphics[width=14cm]{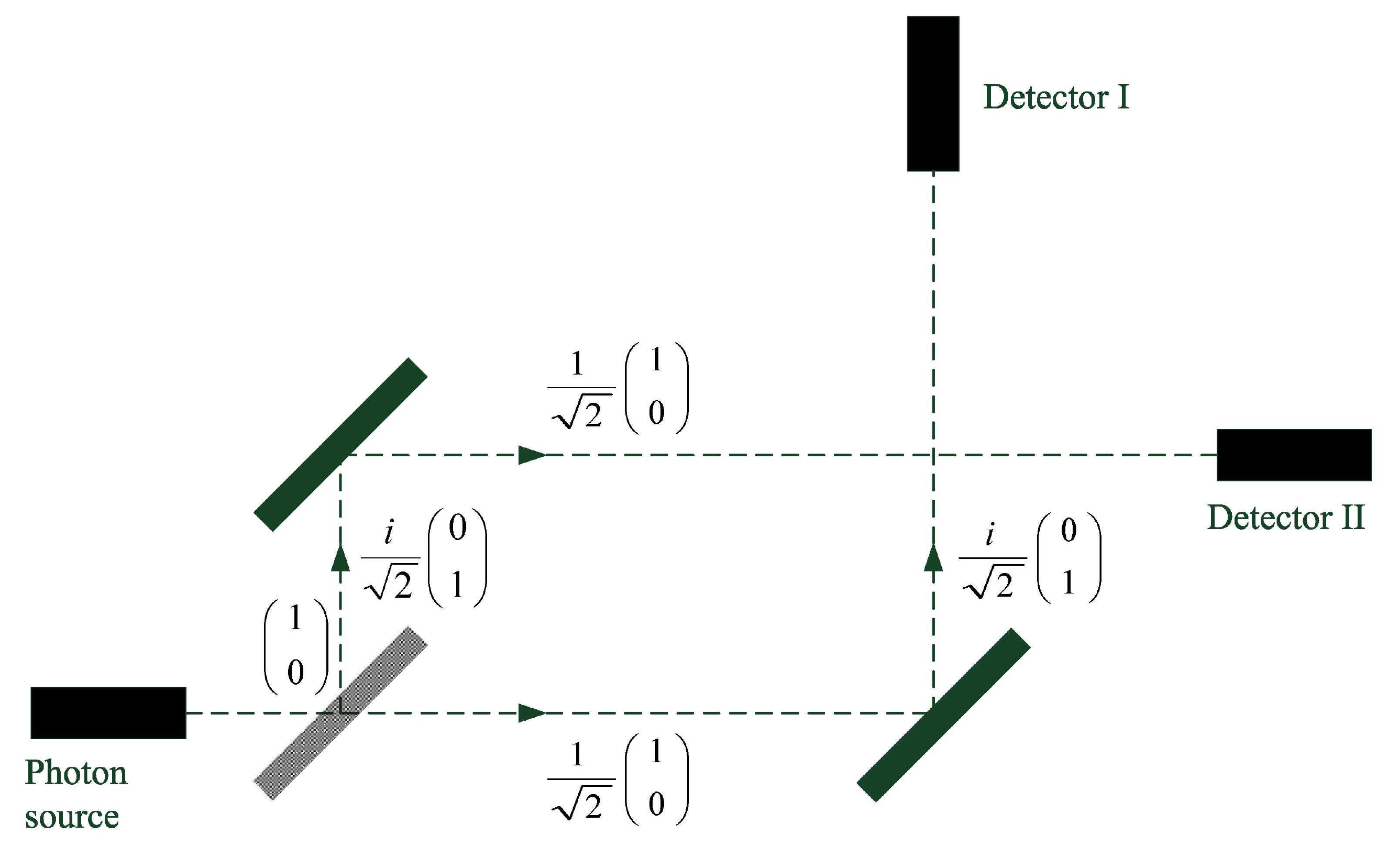}

\begin{Figure} \label{fig2}
The $H$ device with detectors.
\end{Figure}
\end{center}
\end{table}

Secondly, the interferometer (see Figure \ref{fig3a}) is discussed. Using the same arguments, it follows that
\begin{equation} \label{eq5+}
\begin{array}{lcl}
{\displaystyle P_1H RH{1 \choose 0}} & = & {\displaystyle P_1H \left(\frac{i}{\sqrt{2}}{0 \choose 1}- \frac{1}{\sqrt{2}}{1 \choose 0}\right)}\\[0.5cm]
 & = & {\displaystyle P_1 \left(\frac{i}{\sqrt{2}}\left(\frac{i}{\sqrt{2}}{1 \choose 0}+ \frac{1}{\sqrt{2}}{0 \choose 1}\right)- \frac{1}{\sqrt{2}} \left(\frac{1}{\sqrt{2}}{1 \choose 0}+ \frac{i}{\sqrt{2}}{0 \choose 1}\right)\right)}\\[0.5cm]
 & = & {\displaystyle P_1 \left(-\frac{1}{2}{1 \choose 0}+ \frac{i}{2}{0 \choose 1} -\frac{1}{2}{1 \choose 0}- \frac{i}{2}{0 \choose 1}  \right)}\\[0.5cm]
 & = & {\displaystyle P_1 \left(-{1 \choose 0}\right) = 0},
\end{array}
\end{equation}
and
\begin{equation} \label{eq6+}
P_2 HRH {1 \choose 0} = P_2\left(-{1 \choose 0}\right) = -{1 \choose 0}.
\end{equation}
Hence, if a particle will pass the interferometer, it will be observed with probability $|-1|^2=1$ by detector II and with probability zero by detector I.

\begin{table}
\begin{center}
\includegraphics[width=14cm]{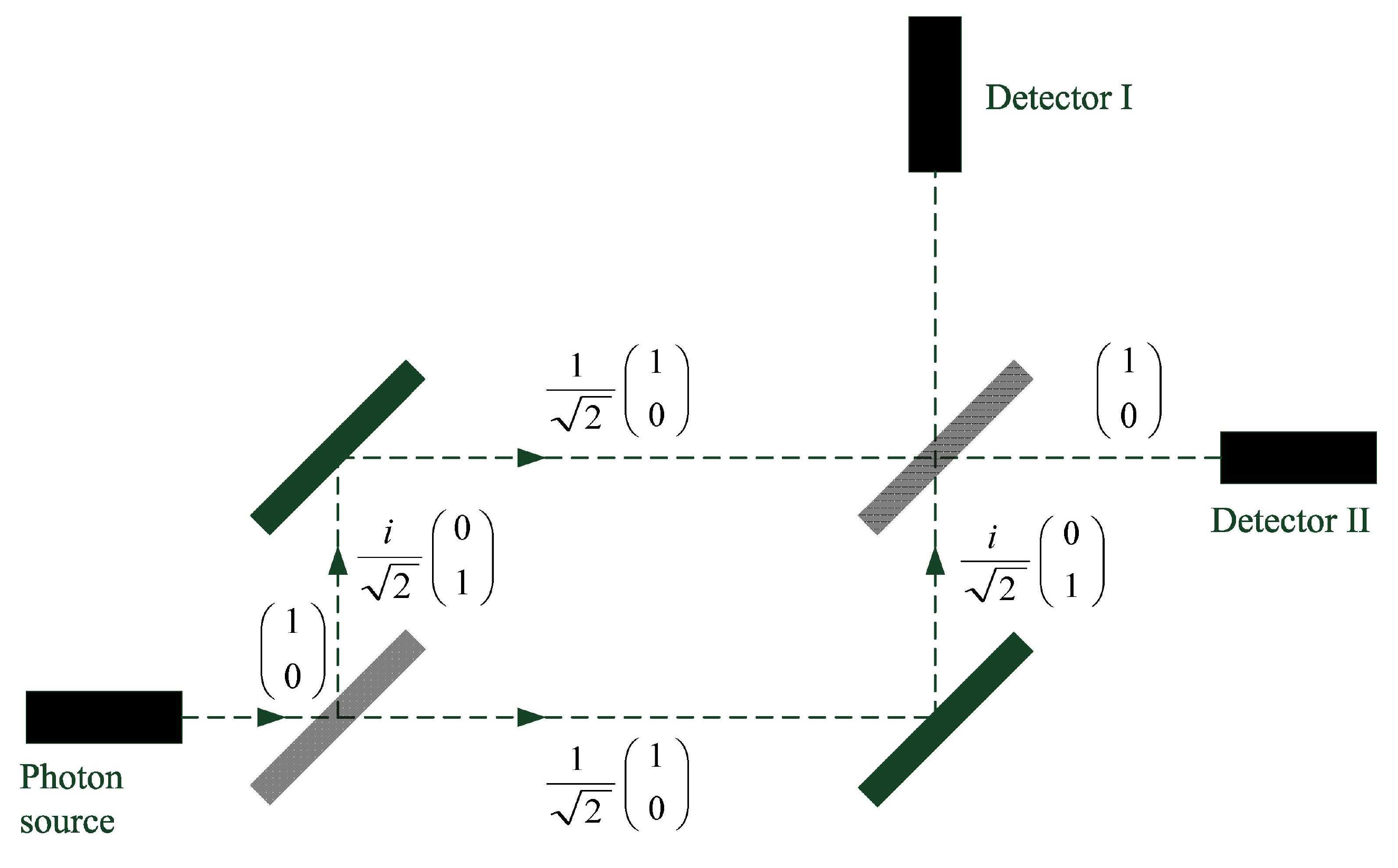}

\begin{Figure} \label{fig3a}
The interferometer.
\end{Figure}
\end{center}
\end{table}



Finally, we discuss interaction-free measurements: A third detector is positioned in the upper path of the interferometer. Then
\begin{equation} \label{eq7+}
P_3 H {1 \choose 0} = P_3 \left(\frac{1}{\sqrt{2}}{1 \choose 0}+ \frac{i}{\sqrt{2}}{0 \choose 1}\right) =  \frac{i}{\sqrt{2}}{0 \choose 1},
\end{equation}
and with probability $\frac{1}{2}$ the particle will be detected by the third detector. In terms of Elitzur and Vaidman the bomb explodes. With the same probability the particle will move along the lower path, and it follows that
\begin{equation} \label{eq8+}
\begin{array}{lcl}
{\displaystyle P_1 HR\left( \frac{1}{\sqrt{2}}{1 \choose 0}\right)} & = & {\displaystyle P_1H \left(\frac{i}{\sqrt{2}}{0 \choose 1}\right)}\\[0.5cm]
 & = & {\displaystyle P_1 \left(-\frac{1}{2}{1 \choose 0}+ \frac{i}{2}{0 \choose 1}\right)}\\[0.5cm]
 & = & {\displaystyle \frac{i}{2}{0 \choose 1}},
\end{array}
\end{equation}
and
\begin{equation} \label{eq9+}
P_2 H R \left(\frac{1}{\sqrt{2}}{1 \choose 0}\right) = -\frac{1}{2}{1 \choose 0}.
\end{equation}
Hence, with probability $\frac{1}{4}$ the particle will be detected in the detectors I and II. Nothing spectacular happened, interaction-free measurement is the immediate consequence of simple postulates.

It is important to notice that the above results are obtained with classical logic and without using coherent wave-like properties. Describing the single devices as linear transformations applied to probability states in an ordered universe suffice. Only these weak assumptions (compared, for example, to many worlds assumptions or to the Copenhagen interpretation) are needed to prove that the paradoxes due to measurements and the measurement problem itself have vanished. The explanation of the above measurements is imperative when using linearity in a higher-dimensional space only. Notice that linear operations provide the most simple rules, are in the sense of Einstein, and are better than to cause schizophrenic particles.

The linear transformations of probability states contain the postulates due to Feynman when using  appropriately inner products for computing the probability amplitude:

(i) The probability  of a quantum event is given
by the square $|\psi|^2$ of the absolute value of
a complex number $\psi$ called the probability
amplitude of the event.

(ii) When an event can take place in n different
ways, the probability amplitude is proportional
to the sum of the
probability amplitudes of each way considered
separately: $\psi = c(\psi_1 + \psi_2 + \cdot + \psi_n)$,
where $c$ is the normalization constant.

(iii) The probability amplitude of n independent events is the product
of the single probability amplitudes.

(iv) If an experiment is carried out that can
determine which of the alternative ways has
actually taken place, the interference is lost.

The Feynman rules are often used to explain the double-slit experiment. But these rules as well as the previous postulates are mathematical descriptions only. There is no ontological justification.
It is the purpose of the next section to provide such a justification. In the following, I use the notion of ontology in order to distinguish desirable and convenient mathematical concepts from physical objects and physical experience.

Other experiments (polarization of photons, Stern-Gerlach devices, e.t.c) can be described and ontologically considered in a similar manner.

\section{Further Consequences}

\subsection{Time}
The basic reason for many paradoxes in physics is the understanding of time. The present time is defined in terms of comparing us with periodical procedures, say a pendulum clock or a light clock. Then it follows that velocity or momentum can be computed in terms of the four variables position and time leading to classical mechanics. The way from classical mechanics to relativity theory provides Einstein's fundamental postulate that space and time are linked together in a way that, independently of the reference frame, the speed of light is always constant. This  seems to be paradoxical when compared to our daily experience: I move at very high speed in the same direction as a photon, and no matter how fast I am, the photon will always move from me at the same speed. {\it It follows that the photon can choose his velocity independently of any motions in the space-time, and its velocity is not linked to space as other particles.} Hence, it has three independent parameters: the direction and the length (the maximal speed differs in matter). Since the photon moves in the three-dimensional space of positions, we obtain six independent parameters. Hence, {\it photons seem to be ontological objects of an at least six-dimensional universe.}

Einstein's fundamental postulate implies the
relativity of motion together with the fact that there can be no absolute frame in the present four-dimensional universe. Looking more carefully, it is not clear what is past and what is future in relativity theory. A photon  is an object of our present universe as we are, but its Lorentz factor is undefined and its time does not pass. Hence, photons should have no past and also no future. They are particles that carry the information of one moment in form of points as on a photo. On the other hand, each photon has a well-defined past and future; it was generated and possibly it interacts thermodynamically with my video camera. Hence, they are also ontological objects of an absolute time ordering, other than the relative time.

One problem in physics is the astonishing fact that the physical laws (except the second law of thermodynamics) are reversible, which is converse to our experience that almost nothing is reversible. This reversibility in the laws is a consequence of the relative time that has no future and no past, yielding a second argument of a necessary absolute time ordering.
This ordering is closely related to the second law of thermodynamics: Entropy cannot decrease, and can be thought of as a kind of second clock; it is a state function w.r.t. $T$ that accounts for the effects of irreversibility.

Another argument for a six-dimensional ordered universe is the obvious fact of our freedom.  A taxi-driver always determines his position and velocity without permanently looking at his clock. His degree of freedom is six, at least when looking at future decisions. He decides where the taxi will go and not his clock. {\it The ordered universe $(X,V,T)$ provides general rules for the present and the future}. The absolute time is in principle completely independent of the relative  time. In the present universe the  relative time, position, momentum, mass, distance, e.t.c. are observables that only have a relative meaning. 

Photons have (i) zero rest mass, (ii) they have a velocity vector in the direction of motion with maximal speed $c$,  and (iii) they carry information about fixed positions $x$ (dimension 3) of the past, but their position is not defined  in the framework of relativity theory.  They can be represented by the quantity $(\inf,c,T)$ with the above interpretation.

Gravitons are hypothetical elementary particles responsible for the force of gravity.
They are postulated in quantum field theory. They have (i) zero motion mass, no velocity vector, are fixed in position, and (ii) they have an acceleration vector $g$ expressing gravitational forces.  They can be represented by the quantity $(x,\inf,T)$ with the above meaning.
From an complementary  point of view (w.r.t. the photons), they (iii) should carry information about possible velocities (dimension 3) that can be expressed in terms of future photos.  Both types of particles are embedded in a six-dimensional linear space of points and vectors.
They justify the postulates 4 and 5, expressing the close-by idea that in a linear space linear operations are the natural ones.

The relative time and our four-dimensional curved space-time rob, not surprisingly, two degrees of freedom, since possibilities of the future must be replaced by realizations that are caused by time-entanglements. The information about the realizations in our present universe immediately disappears to a large extent. Only measured information is maintained. If a particle has passed a double slit and was not measured there, then this information is lost forever. We can only say that the particle is on the interference pattern and it has passed one of the slits. The best available information is described in terms of probability waves: All statements like ``the particle behaves like a wave'' or  ``the particle passes simultaneously all paths'' are correct when formulated for the past. The wave function only represents an observer's knowledge about the past. From this point of view the Copenhagen interpretation
and Born's rule describe the past. With respect to the past, it seems that a particle is a wave and collapses in the present. Unitary transformations express the loss of information which suddenly disappears (collapses) iff the information is measured in the present. I think the deeper reason of many paradoxes due to the measurement in theory and practice is that all information which we obtain is from the past, and thus has wave-like properties. The problems and paradoxes due to the measurement problem do not occur in the ordered universe. Quantum logic is the logic for describing past events.

\subsection{Decoherence}
Decoherence is described as a phenomenon which occurs when a particle or a quantum system interacts with its environment: Then due to a thermodynamically irreversible way  the quantum superposition of the system and the environment's probability wave can no longer interfere with each other. As a consequence, the considered system together with the environment  behaves like a classical statistical ensemble of the different elements rather than like a single coherent quantum superposition. Decoherence has been a subject of active research for the last two decades.

In the GMc-interpretation coherence and decoherence can be explained in terms of the past, the present and the future. The probability states of the future describe decoherence. The realization of the present are the classical states. These are then transformed to coherent superpositions representing the loss of knowledge of all details that were not measured in the present. Therefore, the arrangement of events is to look from future possibilities (described by a decoherent probability state) to the present (described by classical states), and then to the past (described by coherent superpositions).

\subsection{Complementarity and Uncertainty}
Complementarity is a basic principle of quantum mechanics introduced
by Bohr saying that  all properties of physical quantities exist
in pairs, which are called complementary or conjugate pairs.
It refers to effects such as the wave-particle dualism, where different measurements expose particle-like or wave-like properties. Similar complementary properties appear by measuring the spin or the polarization of photons. This concept is considered to be one of the most distinguishing and sophisticated characteristics of quantum mechanics. A
profound aspect and a consequence of complementarity are the uncertainty principles. For
example, a particle can be measured with greater accuracy of its
position only in even trade for a loss in accuracy of measuring its
momentum. The ultimate limitations of the precision in measurements
are quantified by Planck units.
Complementarity and uncertainty dictate that properties and actions
in the physical world are therefore non-deterministic to some
degree. Mathematically, this concept corresponds to non-commuting Hermitian operators, and captures the phenomenon where two quantities do not have simultaneously sharp observable values.

Not surprisingly, there are very different interpretations in
literature.
The uncertainty principle is often explained as the statement that the measurement of position necessarily disturbs a particle's momentum and vice versa. Hence, this principle is an observer effect.
In a modern context of quantum mechanics, classical states with both, definite position and momentum, do not exist, implying that not the measurement apparatus is at fault.
Heisenberg did not focus on the mathematics, he tried to establish that uncertainty is actually a property of the world, and he used physical arguments based on Planck's constant $\hbar$, and tried to avoid the full quantum mechanical formalism.

Not obvious is the energy-time uncertainty principle. Energy
is related to time in the same way as momentum does to position in space. This principle was clear but difficult to many
early founders, because the time at which the
particle has a given state is not an operator belonging to the
particle, it is a parameter describing the evolution of the system.
Landau joked ``To violate the time-energy uncertainty relation all I
have to do is measure the energy very precisely and then look at my
watch''. Einstein and Bohr understood the heuristic meaning of
the principle as follows: A state which only exists for a short time cannot
have a definite energy. In order to have a definite energy, the
frequency of the state needs to be accurately defined, and this
requires the state to hang around for many cycles, the reciprocal of
the required accuracy.

{\it In the GMc-interpretation all quantities have at each absolute ordering time sharp observable classical values, only the knowledge may not be available.}  Complementarity is already defined as the concept of entangled variables that supplement to states. Position and velocity, for example, are entangled by time. Evidently both properties cannot be measured at the same ordering time. It makes a difference if one entangled property is measured before the other, because the measured property is in the past when the other one will be measured. Therefore, the reason of the uncertainty principles is not the mathematical fact that operators do not commute; non-commuting observables are a consequence of entanglements in the present, and the reason is simply the transition from the present  to the past. Energy and relative time are complementary, since these two quantities determine the classical states via Hamilton's equations, and the relative time is measurable as any other observable in the present.

\subsection{Locality and Realism}
Realism, in the context of classical physics, is the assumption that properties of particles and systems must objectively have pre-existing observable values for any possible measurement.
Einstein liked to say that the Moon is "out there" even when no one is observing it. As already postulated and mentioned, in the GMc-interpretation a particle in the present has sharp observable values, and hence realism is preserved.

The principle of locality means that distant objects cannot affect on one another. In other words, an object is only influenced directly by its immediate surroundings.
Sometimes the measurement can be performed from great distance. If, for example, two photons are emitted in opposite directions from the decay of positronium, the momentum of the two photons is opposite, i.e. they are in a maximally entangled Bell state. By measuring the momentum of one particle, the momentum of the other is determined.

In many world interpretations realism and locality are retained in one world, but they are rejected by the  existent parallel worlds. In the GMc-interpretation locality must be viewed w.r.t. the absolute ordering. In the future the meaning is: There is an apparatus that can generate Bell-states with corresponding probabilities.
Then, in our  present universe one of the classical entangled states of the Bell-state is generated. It is not known which one, but the momentum of one photon is the opposite of the other one, i.e. we have  classical probabilities.  A measurement requires time, hence the generation of the Bell-states is then in the past, and when we measure the momentum of one photon, then instantaneously and not surprisingly we obtain the information about the momentum of the other photon.

Bell proved, by assuming local realism and by a clever argument based on classical probability, that correlations between measurements are bounded in a way of an inequality, the well-known
Bell inequality. This inequality implies that quantum mechanics or local realism is wrong.
Bell test experiments overwhelmingly show empirical evidence against local realism and in favor of QM. But it is known, however, that there may be experimental problems and several sources of errors that affect the validity of the results. The term ``loophole'' is frequently used to denote these problems. As already mentioned, in the GMc-interpretation, measurement is a problem of time-entanglement and hence may also lead to a loophole.

\begin{table}
\begin{center}
\includegraphics[height=7cm]{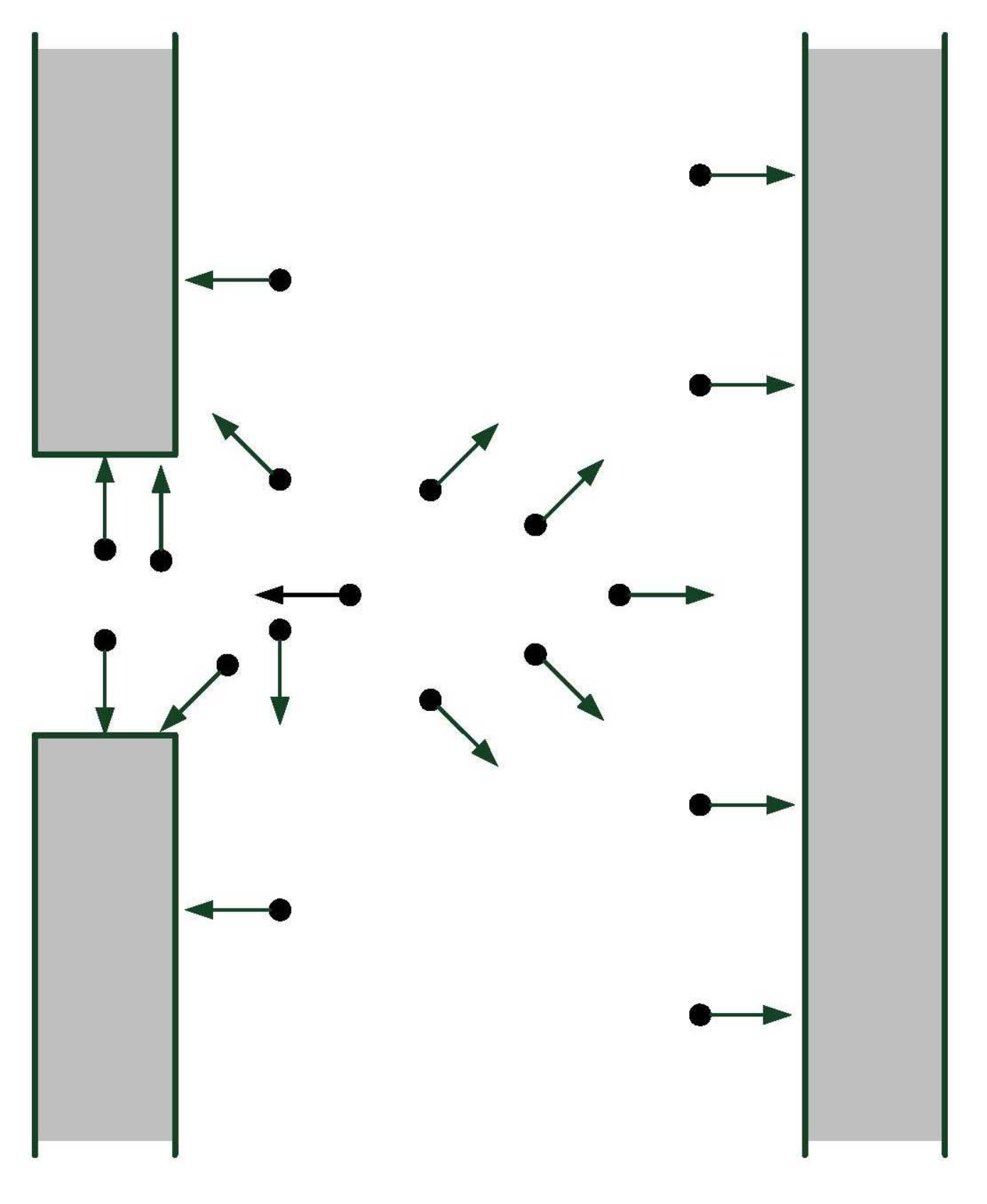}
\begin{Figure} \label{oneslit}
One-Slit.
\end{Figure}
\end{center}
\end{table}

\subsection{Realism of the Future}
Realism, in a broader philosophical sense, is a subject of ontology, the study of conceptions of reality and the nature of being. The basic question is:``What exists and why ?'' Philosophers give different answers to this question.  In the philosophy of metaphysical realism, the physical reality is completely ontologically independent of our conceptual schemes, linguistic practices, mathematics, beliefs, etc. In Physics the entities and nouns of ontology  apply to particles such as electrons or photons, but also to mass and time. If thinking in this way one must answer the question, which ontological concepts are responsible for interference patterns, and more general for postulates in physics.

Photons, gravitons,  and the degree of freedom were already discussed as entities for the formulated postulates. Now, some arguments are presented that should justify interference pattern as entities of the future, that is {\it interference patterns exist and are known when the apparatus is built, before any particle passes the experiment.}

\begin{table}
\begin{center}
\includegraphics[height=15cm]{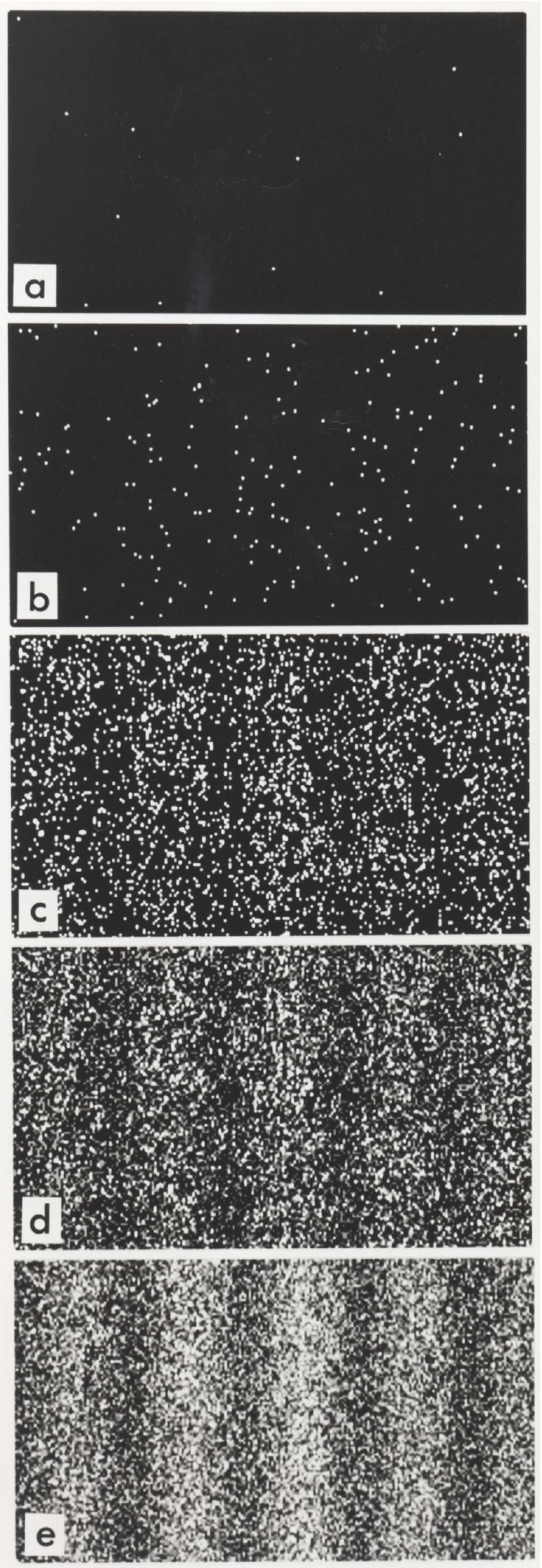}

\begin{Figure} \label{doubleslit}
Image Double-Slit experiment \cite{Tonomura}.
\end{Figure}
\end{center}
\end{table}

Let us look at the key experiment, the double slit. If only one slit is
open the whole apparatus and environment is in motion and moves through the gravitons. The gravitons direct there gravitational vectors according to the mass-distribution of the apparatus, similar like in Figure \ref{oneslit}. This vector field produces the well-known interference pattern, and thus exists before a particle is in the experiment.
If the slit becomes smaller, then position becomes more precise, and the gravitational forces close to the slit increase. Hence, when a particle will be put into experiment, the uncertainty of momentum increases. We have measured the momentum of the particle later than it's position, yielding the uncertainty.
If both slits are open, then the gravitational field changes due the mass between the two slits yielding a typical  interference pattern like in Figure \ref{doubleslit}.

Looking at this figure, my point of view is that it is like an oracle or a photo that evidences the future, not the present and not the past. It is more obvious to make an apparatus consisting of  about $10^{25}$ atoms responsible for interference than one particle or than splitting the worlds.

\section{Conclusion}
I am aware of the fact that this interpretation yields a not familiar conception of motion and time. However, it avoids many worlds interpretation and  its determinism that all parts of a decision are realized. It is in the sense of Ockham, it can be understand in terms of classical mechanics, and, what is most important to me, provides freedom of decisions.

\section{Acknowledgement} I would like to thank Ronald Mirman for his comments on my first paper about the GMc-interpretation, and for pointing out to me the importance of our four-dimensional present universe.


\vspace{4cm}
I have few references, because I only used well-known topics in this manuscript that have appeared in very many places and are numerously referenced therein. It is impossible to list all of them.


\bibliography{H:/bib/extern,H:/bib/ti3}

\end{document}